\documentstyle[aps,amstex,amssymb,graphicx,preprint]{revtex}

\begin{document}
\title{Limitations of the two-media approach in calculating
magneto--optical properties of layered systems}
\author{A. Vernes}
\address{Center for Computational Materials Science, \\
Technical University Vienna, Gumpendorferstr. 1a, A--1060 Vienna, Austria}
\author{L. Szunyogh}
\address{Center for Computational Materials Science, \\
Technical University Vienna, Gumpendorferstr. 1a, A--1060 Vienna, Austria\\
Department of Theoretical Physics, Budapest University of Technology and\\
Economics, \\
Budafoki \'{u}t 8, H--1521 Budapest, Hungary}
\author{P. Weinberger}
\address{Center for Computational Materials Science, \\
Technical University Vienna, Gumpendorferstr. 1a, A--1060 Vienna, Austria}
\date{\today}
\maketitle

\begin{abstract}
It is shown that in polar geometry and normal incidence the $2\times 2$
matrix technique -- as discussed in detail in a preceeding paper\cite{VSW02d}
-- accounts correctly for multiple reflections and optical interferences,
and reduces only in the case of a periodic sequence of identical layers to
the Fresnel formula of reflectivity, which in turn is the theoretical basis
of the two--media approach, widely used in the literature to compute
magneto--optical Kerr spectra. As a numerical example ab--initio
calculations of the optical constants for an fcc Pt semi--infinite bulk
using the spin--polarized relativistic screened Korringa--Kohn--Rostoker
method show very good agreement with experimental data.
\end{abstract}


\section*{Introduction}

In the last few years magneto-optics started to become of prime interest in
dealing with magnetic multilayer systems and magnetic nanostructures
(submonolayer coverages of substrates with magnetic material): Kerr
measurements not only turned out to be one of the standard experimental
tools applied, new applications, in particular in the context of
time-resolved techniques, are presently believed to lead to fast
magnetization switching devices. In many cases, however, theoretical
descriptions even for time-integrated magneto-optical effects are lagging
behind the experimental efforts mainly because relevant schemes to deal with
semi-infinite systems -- not to speak of nanostructured materials -- are not
commonly used. Bulk-like approaches (assumed three-dimensional periodicity)
and the so-called two-media model \cite{RS90} (assumed homogeneity) are
still considered to be sufficient to deal with magnetically inhomogeneous
layered systems. 

In a recent paper, \cite{VSW02d} the authors showed that only by including
multiple reflections and optical interferences, e.g.{\it ,} via the $2\times
2$ matrix technique \cite{Man90,Man95}, realistic ab--initio
magneto--optical Kerr spectra for semi--infinite layered systems can be
obtained. It is the purpose of the present paper to prove first analytically
and then numerically that only in the case of periodic layered systems,
i.e., in by definition homogeneous systems, the $2\times 2$ matrix technique
reduces to the two--media approach. Formulated oppositely, this implies that
the two--media approach is strictly valid only for this kind of systems.
This is illustrated for fcc Pt viewed as a periodic layered system, because
Pt frequently serves as substrate for magneto--optically active multilayers. %
\cite{WBG+92,FBT00}

\section*{Theoretical approach}

Assuming polar geometry and the magnetization $\vec{M}_{p}$ in all layers $p$
to point along the ${\rm \ z}$--direction, the layer--resolved permittivity
in cubic, hexagonal or tetragonal systems is given by 
\begin{equation}
{\bf \tilde{\varepsilon}}^{p}=\left( 
\begin{array}{rll}
\tilde{\varepsilon}_{{\rm xx}}^{p} & \tilde{\varepsilon}_{{\rm xy}}^{p} & 0 \\ 
-\tilde{\varepsilon}_{{\rm xy}}^{p} & \tilde{\varepsilon}_{{\rm xx}}^{p} & 0 \\ 
0 & 0 & \tilde{\varepsilon}_{{\rm xx}}^{p}%
\end{array}
\right) \,,\qquad (p=1,\ldots ,N)\,,  \label{eq:formepsp}
\end{equation}
provided that a possible anisotropy in the diagonal elements can be
neglected, i.e., \ $\tilde{\varepsilon}_{{\rm zz}}^{p}\simeq \tilde{%
\varepsilon}_{{\rm xx}}^{p}$. In case of normal incidence, $\tilde{n}_{p{\rm %
x}}=\tilde{n}_{p{\rm y}}=0$, four electromagnetic beams corresponding to the
complex refractive indices 
\[
\left\{ 
\begin{array}{lllllll}
\tilde{n}_{p{\rm z}}^{(3)} & = & -\tilde{n}_{p{\rm z}}^{(1)} & = & \sqrt{%
\tilde{\varepsilon}_{{\rm -}}^{p}} & \equiv & \tilde{n}_{p{\rm -}} \\ 
\tilde{n}_{p{\rm z}}^{\left( 4\right) } & = & -\tilde{n}_{p{\rm z}}^{\left(
2\right) } & = & \sqrt{\tilde{\varepsilon}_{{\rm +}}^{p}} & \equiv & 
\tilde{n}_{p{\rm +}}%
\end{array}
\right. \ . 
\]
propagate in layer $p$. By considering harmonic fields 
\begin{equation}
\vec{A}\left( z,t\right) ={\cal \vec{A}}\exp \left[ i\left( \tilde{q}%
z-\omega t\right) \right] \exp \left( -\delta t\right) ={\cal \vec{A}}\exp %
\left[ i\left( \tilde{q}z-\tilde{\omega}t\right) \right] \ ,
\label{eq:alghpw}
\end{equation}
such that $\delta >0$ describes the interaction between the layered system
and its neighborhood, $\tilde{q}=q_{0}\tilde{n}$ is the complex wave vector
with $q_{0}$ being the propagation constant in vacuum, and $\tilde{\omega}%
=\omega -i\delta $, beams 1 and 2 propagate along $-{\rm z}$, beams 3 and 4
along $+{\rm z}$. The surface reflectivity matrix, see in particular the
appendix of Ref.~\onlinecite{VSW02d},{} {} 
\begin{equation}
R_{{\rm surf}}=\left( 
\begin{array}{rr}
\tilde{r}_{{\rm xx}} & \tilde{r}_{{\rm xy}} \\ 
-\tilde{r}_{{\rm xy}} & \tilde{r}_{{\rm xx}}%
\end{array}
\right) \ ,  \label{eq:formRsurf}
\end{equation}
which relates the incident electric field to the reflected one, 
\begin{equation}
R_{{\rm surf}}=\left( B_{{\rm vac}}+{\cal D}_{N}{\cal A}^{-1}\right)
^{-1}\left( B_{{\rm vac}}-{\cal D}_{N}{\cal A}^{-1}\right) \ ,
\label{eq:Rsurf}
\end{equation}
is obtained as defined in Eq.~(37) of Ref.~\onlinecite{VSW02d}, in terms of
reflectivity matrices ${\cal R}_{p}$, see Eq.~(32) of Ref.~%
\onlinecite{VSW02d}{\it ,} 
\begin{equation}
{\cal R}_{p}=\left( {\cal B}_{p}+{\cal D}_{p-1}\right) ^{-1}\left( {\cal B}%
_{p}-{\cal D}_{p-1}\right) \ ,\quad p=1,\ldots ,N\ ,  \label{eq:Rmtx}
\end{equation}
and propagation matrices ${\cal C}_{p},$ see Eq.~(30) of Ref.~%
\onlinecite{VSW02d}, 
\begin{equation}
{\cal C}_{p}=\left( 
\begin{array}{cc}
\exp \left( -iq_{0}\tilde{n}_{p{\rm -}}d_{p}\right) & 0 \\ 
0 & \exp \left( -iq_{0}\tilde{n}_{p{\rm +}}d_{p}\right)%
\end{array}
\right) \ ,  \label{eq:Cmtx}
\end{equation}
where $d_{p}$ refers to the thickness of layer $p$. According to Eqs.~(28),
(32) and (35) of Ref.~\onlinecite{VSW02d}{\it \ }the $2\times 2$ matrices
occurring in Eq. (\ref{eq:Rsurf}) are defined as follows

\begin{equation}
{\cal D}_{p}={\cal B}_{p}\left[ {\cal C}_{p}-\left( {\cal C}_{p}\right) ^{-1}%
{\cal R}_{p}\right] \left[ {\cal C}_{p}+\left( {\cal C}_{p}\right) ^{-1}%
{\cal R}_{p}\right] ^{-1}\ ,  \label{eq:Dmtx}
\end{equation}
\begin{equation}
{\cal A}=\left( 
\begin{array}{ll}
1 & i \\ 
i & 1%
\end{array}
\right) \ ,\quad {\cal B}_{p}=\left( 
\begin{array}{rr}
i\tilde{n}_{p{\rm -}} & \tilde{n}_{p{\rm +}} \\ 
-\tilde{n}_{p{\rm -}} & -i\tilde{n}_{p{\rm +}}%
\end{array}
\right) \ ,\quad B_{{\rm vac}}=\left( 
\begin{array}{rr}
0 & 1 \\ 
-1 & 0%
\end{array}
\right) \ .  \label{eq:ABmtx}
\end{equation}
${\cal R}_{N}$ results then recursively starting from the substrate
reflectivity matrix ${\cal R}_{0}=0$. It has been shown{\it \ }in quite some
detail in the appendix of Ref.~\onlinecite{VSW02d} that the layer--resolved
reflectivity matrices ${\cal R}_{p}$ are all diagonal, 
\[
{\cal R}_{p}=\left( 
\begin{array}{ll}
\tilde{r}_{p{\rm -}} & 0 \\ 
0 & \tilde{r}_{p{\rm +}}%
\end{array}
\right) \ , 
\]
where $\tilde{r}_{p{\rm \pm }}$ is the complex reflectivity coefficient of
the right-- and left--handed light in layer $p$. This implies that the
right-- and left--handed circularly polarized components of the incident
linearly polarized light, once they arrived at the surface layer, propagate
independently within the system such that after the first reflection, they
become immediately elliptically polarized.

\section*{Special case of homogeneous systems}

In principle in a finite periodic layered system ($N$ can be large but
finite; for matters of simplicity a simple parent lattice \cite{pw-pmb} is
assumed) all layers have identical layer--resolved permittivities. This
implies in turn that also all matrices ${\cal B}_{p}$ in Eq.\ (\ref{eq:ABmtx}%
) are identical, i.e., are of the form 
\begin{equation}
{\cal B}=\left( 
\begin{array}{rr}
i\tilde{n}_{{\rm -}} & \tilde{n}_{{\rm +}} \\ 
-\tilde{n}_{{\rm -}} & -i\tilde{n}_{{\rm +}}%
\end{array}
\right) \ ,  \label{eq:Bmtx4pls}
\end{equation}
with $\tilde{n}_{{\rm \pm }}=\sqrt{\tilde{\varepsilon}_{{\rm \pm }}}$. The
recursion relation in Eq.\ (\ref{eq:Rmtx}) reduces therefore to 
\begin{equation}
{\cal R}_{p}=\left( {\cal I}+{\cal G}_{p-1}\right) ^{-1}\left( {\cal I}-%
{\cal G}_{p-1}\right) \ ,\quad p=1,\ldots ,N\ ,  \label{eq:Rmtx4pls}
\end{equation}
where ${\cal I}$ denotes the $2\times 2$ identity matrix and 
\begin{equation}
{\cal G}_{p-1}=\left[ {\cal C}_{p-1}-\left( {\cal C}_{p-1}\right) ^{-1}{\cal %
R}_{p-1}\right] \left[ {\cal C}_{p-1}+\left( {\cal C}_{p-1}\right) ^{-1}%
{\cal R}_{p-1}\right] ^{-1}\ .  \label{eq:Gmtx}
\end{equation}

This finite periodic layered structure has to be properly matched to a
semi--infinite system (substrate) of the same material. Inserting ${\cal R}%
_{0}=0$ into Eq.\ (\ref{eq:Gmtx}) for $p=1$, yields ${\cal G}_{0}={\cal I}$,
which substituted into Eq.\ (\ref{eq:Rmtx4pls}), immediately proves that
also ${\cal R}_{1}=0$, and so on. In conclusion, for a periodic layered
system, 
\begin{equation}
{\cal R}_{p}=0,\qquad \forall p=1,\ldots ,N\ .  \label{eq:Rmtx4bulk}
\end{equation}
>From an optical point of view a periodic layered system behaves like \ a
system with no boundaries. Viewed oppositely, Eq.\ (\ref{eq:Rmtx4bulk})
shows that there is a boundary in-between two adjacent layers, if and only
if, the respective layer--resolved permittivities differ.

>From ${\cal R}_{N}=0$ follows immediately that ${\cal D}_{N}={\cal B}$ with $%
{\cal B}$ being defined in Eq.\ (\ref{eq:Bmtx4pls}); hence Eq.\ (\ref%
{eq:Rsurf}) directly yields the surface reflectivity matrix 
\begin{equation}
R_{{\rm surf}}={{\frac{1}{\left( \tilde{n}_{{\rm +}}+1\right) \left( 
\tilde{n}_{{\rm -}}+1\right) }}}\left( 
\begin{array}{rr}
1-\tilde{n}_{{\rm +}}\tilde{n}_{{\rm -}} & -i\left( \tilde{n}_{{\rm +}}-%
\tilde{n}_{{\rm -}}\right) \\ 
i\left( \tilde{n}_{{\rm +}}-\tilde{n}_{{\rm -}}\right) & 1-\tilde{n}_{{\rm +}%
}\tilde{n}_{{\rm -}}%
\end{array}
\right) \ ,  \label{eq:Rsurf4pls}
\end{equation}
i.e., \ $\tilde{r}_{{\rm xx}}$ and $\tilde{r}_{{\rm xy}}$ in Eq. (\ref%
{eq:formRsurf}) assume the following values 
\[
\tilde{r}_{{\rm xx}}=-{{\frac{\tilde{n}_{{\rm +}}\tilde{n}_{{\rm -}}-1}{%
\left( \tilde{n}_{{\rm +}}+1\right) \left( \tilde{n}_{{\rm -}}+1\right) }%
\quad ,\quad \tilde{r}_{{\rm xy}}=-i{{\frac{\tilde{n}_{{\rm +}}-\tilde{n}_{%
{\rm -}}}{\left( \tilde{n}_{{\rm +}}+1\right) \left( \tilde{n}_{{\rm -}%
}+1\right) }}}\quad .}} 
\]
In the case of periodic layered systems the complex reflectivity coefficient
of the right-- and left--handed polarized light is therefore given by 
\begin{equation}
\tilde{r}_{{\rm \pm }}=\tilde{r}_{{\rm xx}}\mp i\tilde{r}_{{\rm xy}}=-{{%
\frac{\tilde{n}_{{\rm \pm }}-1}{\tilde{n}_{{\rm \pm }}+1}}}\ ,
\label{eq:fresnel}
\end{equation}
a relation, which is known in the literature as the Fresnel formula for $s$
--polarization and normal incidence. \cite{Jac75} Eq.\ (\ref{eq:fresnel})
then leads directly to the well--known formula for the complex Kerr angle in
the two--media approach, \cite{RS90} 
\[
\theta _{{\rm K}}+i\epsilon _{{\rm K}}\simeq {{\frac{\tilde{\sigma}_{{\rm xy}%
}}{\tilde{\sigma}_{{\rm xx}}}}}\left( \ 1+{{\frac{4\pi i}{\tilde{\omega}}}}%
\tilde{\sigma}_{{\rm xx}}\right) ^{-1/2}, 
\]
where $\theta _{{\rm K}}$ is the Kerr rotation angle and $\epsilon _{{\rm K}%
} $ the Kerr ellipticity.

Clearly enough in the case of (homogeneous) periodic layered systems the $%
2\times 2$ matrix technique and the two--media approach provide identical
Kerr spectra. However, by using the two--media approach for calculating Kerr
spectra of inhomogeneous layered systems, such spectra are governed almost
exclusively by contributions from the surface layer.

\section*{Application to fcc Pt}

\begin{figure}[tbph]\centering
\includegraphics[width=0.65\columnwidth,clip]{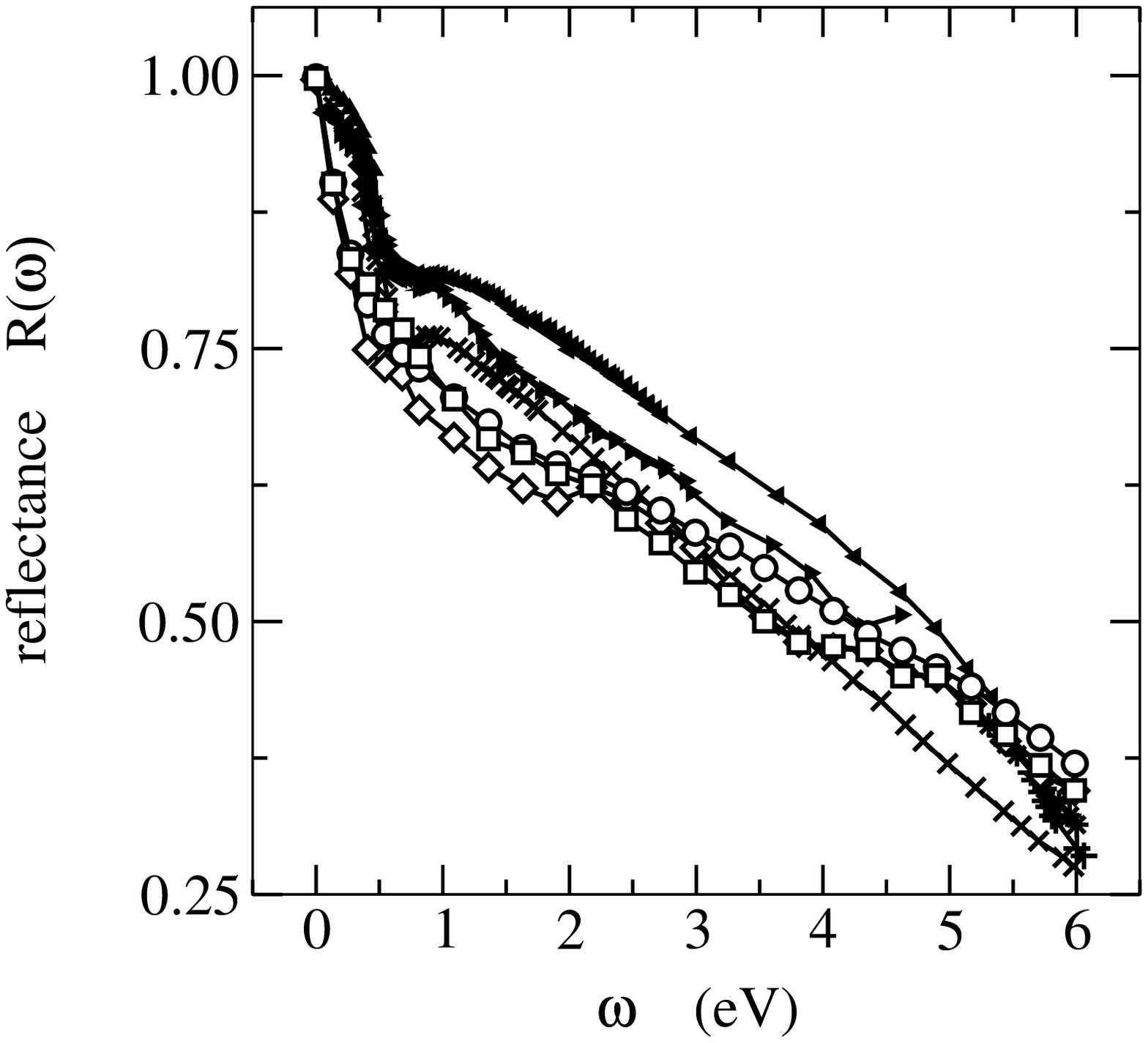}
\caption[Reflectance of fcc Pt bulk for different surfaces.]{Reflectance of
fcc Pt bulk as calculated for the (100), (110) and (111) surface orientation
(diamonds, circles and squares). The experimental data are taken from Refs.\ 
\onlinecite{Wea75} (crosses), \onlinecite{HAH79} (pluses), \onlinecite{SR72}
(stars), \onlinecite{KNN72} (triangle right), \onlinecite{DH64} (triangle
left) and \onlinecite{JPT72} (triangle up).}
\label{fig:ref4ni-ptb12}
\end{figure}

\begin{figure}[tbph]\centering
\includegraphics[width=0.65\columnwidth,clip]{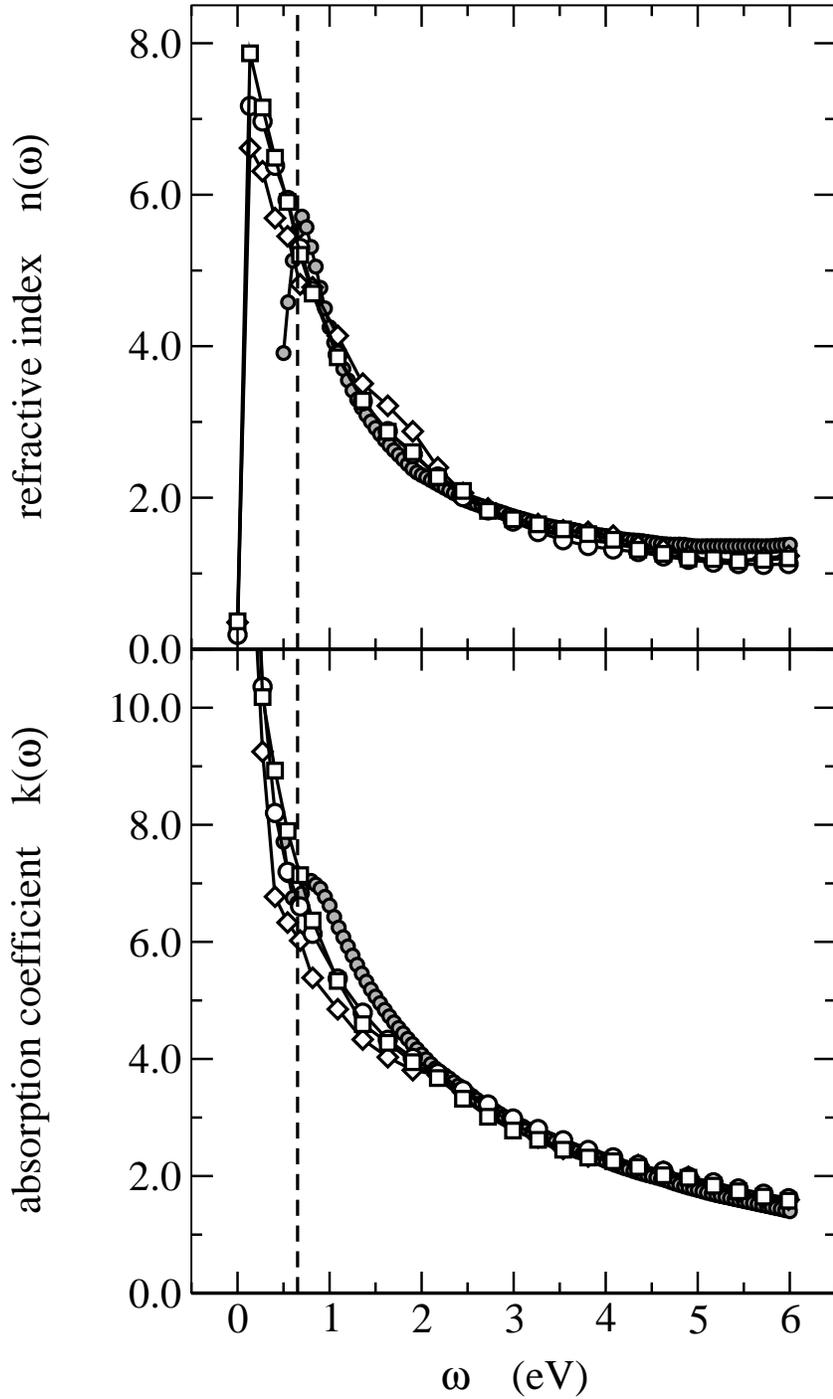}
\caption[Optical constants of fcc Pt bulk for different surfaces.]{Optical
constants of fcc Pt bulk as calculated for the (100), (110) and (111)
surface orientation (diamonds, circles and squares). The experimental data
of Weaver et al. \protect\cite{WKL+81} are displayed as grey circles. The
dotted line marks the photon energy that equals the used lifetime broadening
of $\protect\delta =0.653\ {\rm eV}$.}
\label{fig:optconst-ptb12}
\end{figure}

\begin{figure}[tbph]\centering
\includegraphics[width=0.65\columnwidth,clip]{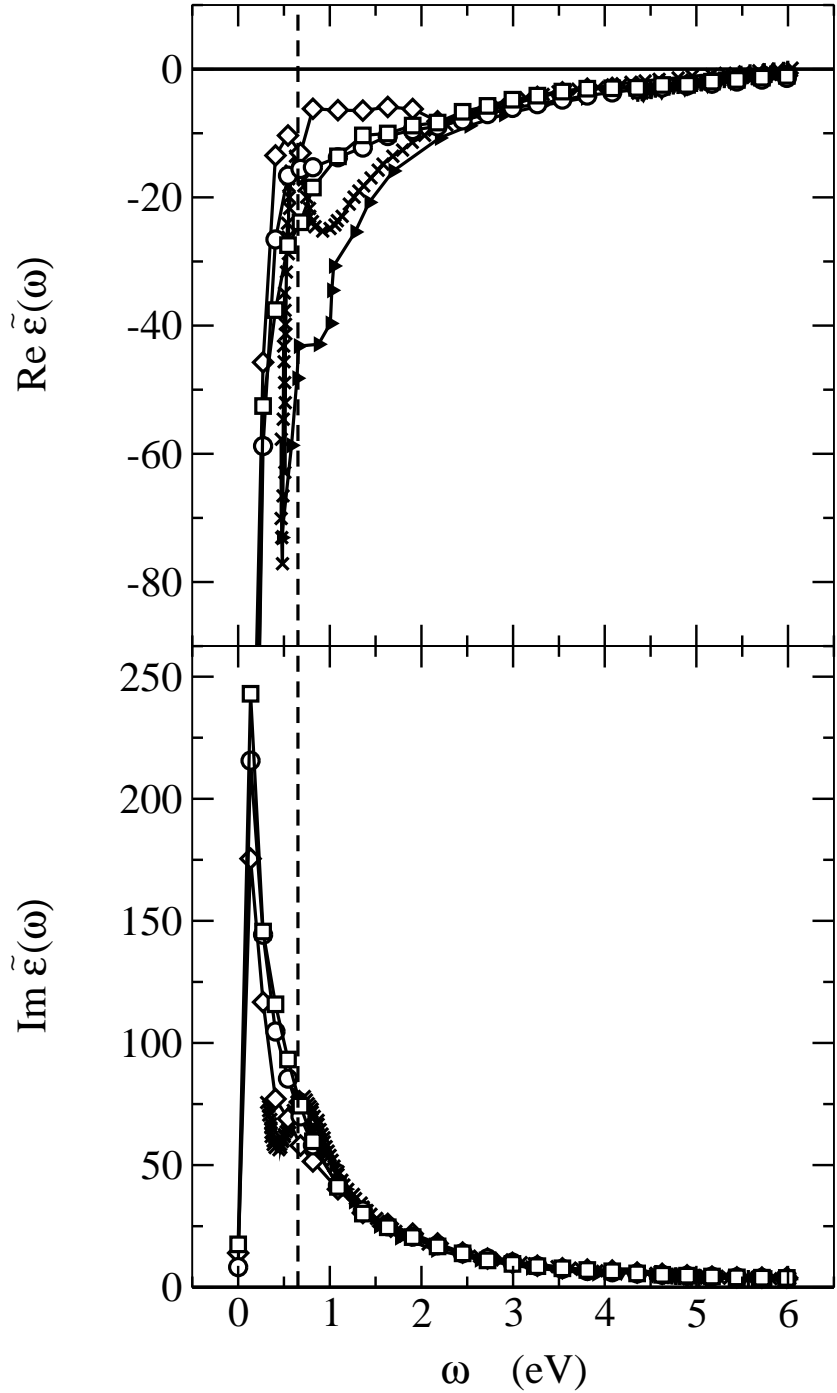}
\caption[Permittivity of fcc Pt bulk for different surfaces.]{Permittivity
of fcc Pt bulk as calculated for the (100), (110) and (111) surface
orientation (diamonds, circles and squares). The experimental data are taken
from Refs.\ \onlinecite{Wea75} (crosses), \onlinecite{HAH79} (pluses), 
\onlinecite{SR72} (stars), \onlinecite{KNN72} (triangle right) and 
\onlinecite{YSH68} (triangle down). The dotted line marks the photon energy
that equals the used lifetime broadening of $\protect\delta =0.653\ {\rm eV}$%
.}
\label{fig:eps-ptb12}
\end{figure}

In Figs.\ \ref{fig:ref4ni-ptb12} -- \ref{fig:eps-ptb12} the optical
constants of fcc Pt bulk as calculated via the $2\times 2$ matrix technique
for normal incidence and different surface orientations are compared with
available experimental data (Ref.\ \onlinecite{WKL+81} and references cited
therein). Because fcc Pt is paramagnetic, the right-- and left--handed
complex reflectivity coefficients are equal, $\tilde{r}_{{\rm +}}\left(
\omega \right) =\tilde{r}_{{\rm -}}\left( \omega \right) \equiv \tilde{r}%
\left( \omega \right) $, and the reflectance $R\left( \omega \right) $ given
in Figs.\ \ref{fig:ref4ni-ptb12} is in fact $\tilde{r}\left( \omega \right)
^{2}$. In accordance with Eq.\ (\ref{eq:alghpw}) the refractive index $%
n\left( \omega \right) $ and the absorption coefficient $k\left( \omega
\right) $ in Fig.\ \ref{fig:optconst-ptb12} simply correspond to the real
and imaginary part of the complex refractive index $\tilde{n}\left( \omega
\right) =\sqrt{\tilde{\varepsilon}\left( \omega \right) }\equiv n\left(
\omega \right) +ik\left( \omega \right) $. The total (complex) permittivity $%
\tilde{\varepsilon}\left( \omega \right) $ displayed in Fig.\ \ref%
{fig:eps-ptb12} is evaluated assuming a periodic layered system consisting
of $N$ Pt layers on top of a semi--infinite Pt substrate, i.e., 
\begin{equation}
\tilde{\varepsilon}\left( \omega \right) =\frac{1}{N}\sum_{p,q=1}^{N}\tilde{%
\varepsilon}^{pq}\left( \omega \right) \ ,  \label{eq:epspq2epstot}
\end{equation}
where 
\begin{equation}
\tilde{\varepsilon}^{pq}\left( \omega \right) =\delta _{pq}+{{\frac{4\pi i}{%
\tilde{\omega}}}}\tilde{\sigma}^{pq}\left( \omega \right) \ ,\quad
p,q=1,\ldots ,N\ ,  \label{eq:sgmpq2epspq}
\end{equation}
with $\tilde{\sigma}^{pq}\left( \omega \right) $ being the inter-- and
intra--layer contributions to the optical conductivity and $\tilde{\omega}%
=\omega -i\delta $. Eqs.\ (\ref{eq:epspq2epstot}) and (\ref{eq:sgmpq2epspq})
follow directly from the Fourier--transformed (linear) material equations %
\cite{AG66} by assuming that each layer $p$ can be viewed as a homogeneous
medium and therefore the layer--resolved optical conductivity $\tilde{\sigma}%
^{p}\left( \omega \right) $ is related to the layer--resolved permittivity
by 
\begin{equation}
\tilde{\varepsilon}^{p}\left( \omega \right) =1+{{\frac{4\pi i}{\tilde{\omega%
}}}}\tilde{\sigma}^{p}\left( \omega \right) \ ,\quad p,q=1,\ldots ,N\ .
\label{eq:sgmp2epsp}
\end{equation}

The $\tilde{\sigma}^{pq}\left( \omega \right) $ in Eq.\ (\ref{eq:sgmpq2epspq}%
) are calculated in terms of Luttinger's formula \cite{Lut67} using the
spin--polarized relativistic screened Korringa--Kohn--Rostoker method (SKKR)
for layered systems \cite{WLB+96,SUWK94,SUW95,USW95}, contour integration
techniques, \cite{VSW02a,SW99} the Konrod quadrature and the cumulative
special points method for the occurring Brillouin zone integrals.\cite{VSW01}
It should be noted that Luttinger's formula uses a vector potential
description of the electric field and has several advantages over the
well--known Kubo formula \cite{Kub57}: both, the absorptive and dissipative
parts of the conductivity tensor are included without using Kramers--Kronig
relations, and so are all inter-- and intra--band contributions avoiding
thus a phenomenological Drude term in order to mimic the latter
contributions.

In the present study of fcc Pt bulk (lattice parameter $7.4137\ {\rm a.u.}$)
all complex energy and $\vec{k}$--space integrals are performed with an
accuracy of $10^{-3}$ (in atomic units) for all surface orientations
considered. The electronic temperature amounts to $T=300\ {\rm K}$.
Furthermore, a lifetime broadening of $0.048\ {\rm Ryd}\simeq 0.653\ {\rm eV}
$ is used and $2$ and $37$ Matsubara poles in the lower and upper
semi--plane, respectively, are taken into account.

As can be seen from Fig.\ \ref{fig:ref4ni-ptb12} the reflectance $R\left(
\omega \right) $ of Pt as calculated for normal incidence and different
surface orientations agrees very well with the experimental data.
Unfortunately the experimental data \cite{WKL+81} are not specified with
respect to surface orientations, i.e., no surface normal is listed. The
dependence of all the calculated optical constants on the surface
orientation reflects, however, merely the fact that in principle in
spectroscopical experiments only physical properties of semi-infinite
systems, i.e., of solid systems with surface, are recorded, see also Figs.\ %
\ref{fig:optconst-ptb12} and \ref{fig:eps-ptb12}.

A close inspection of Fig.\ \ref{fig:ref4ni-ptb12} in the vicinity of very
low photon energies reveals that the fine structure around 0.5 eV seen in
the experimental reflectance is missing in the theoretical curves although
the maximum in the refractive index as a function of the photon energy is
well reproduced. This discrepancy between theory and experiments is caused
by an intrinsic feature of the applied approach: the finite lifetime
broadening $\delta $ that enters Luttinger's formula smears out possible
fine structures for $\omega \leq \delta $ (the value of $\omega =\delta $ is
indicated in Figs.\ \ref{fig:optconst-ptb12} and \ref{fig:eps-ptb12} as a
dotted line).

In Figs.\ \ref{fig:ref4ni-ptb12} -- \ref{fig:eps-ptb12} the theoretical data
refer to a total of $N=12$ Pt layers, because from calculations for $%
N=3,\ldots ,18$ Pt layers (not shown in here) no significant $N$ dependence
of optical constants was found: the optical constants obtained for $N=3$ are
almost as good as those obtained with $N=18$ Pt layers in the system.

\begin{figure}[tbph]\centering
\includegraphics[width=0.65\columnwidth,clip]{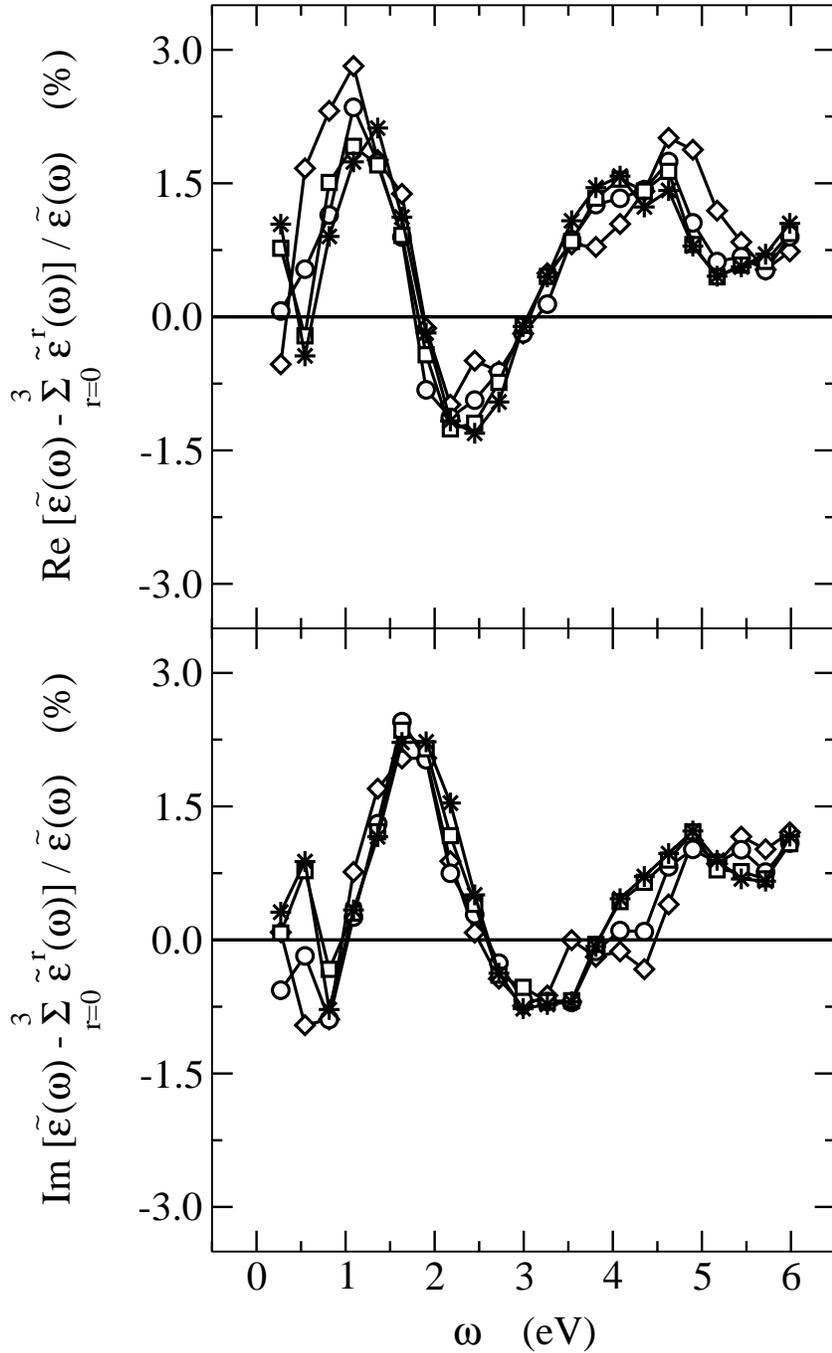}
\caption[Relative error for fcc Pt(111): $N$--dependence.]{Relative error
made by approximating the permittivity of fcc Pt$_{N}$$\mid $Pt(111) by the
sum over $\tilde{\protect\epsilon}^{\,r}\left( \protect\omega \right) $
permittivities with $r\equiv \mid p-q\mid =0,\ldots ,3$, see text. Up and
down triangles, diamonds, circles, squares and stars denote data obtained
for $N=3,6,9,12,15,18$.}
\label{fig:err-ptb111}
\end{figure}

An extensive analysis of the inter-- and intra--layer contributions to the
permittivity $\tilde{\varepsilon}^{pq}\left( \omega \right) $ showed that in
the case of periodic layered systems (simple lattices) the $pq$-like
contributions to the permittivity are only functions of the relative
position of layers $p$ and $q$, i.e.,\ $\tilde{\varepsilon}^{pq}\left(
\omega \right) =\tilde{\varepsilon}^{\left| p-q\right| }\left( \omega
\right) $; the layer--resolved permittivities $\tilde{\varepsilon}^{p}\left(
\omega \right) $ are dominated by the corresponding intra--layer
contribution $\tilde{\varepsilon}^{pp}\left( \omega \right) $. Thus is not
surprising that in the case of periodic layered systems, the following
approximative form applies 
\begin{equation}
\tilde{\varepsilon}\left( \omega \right) \simeq \frac{1}{N}\sum_{\left|
p-q\right| =0}^{3}\tilde{\varepsilon}^{\left| p-q\right| }\left( \omega
\right) \,,  \label{eq:esptotapprox}
\end{equation}
see Fig.\ \ref{fig:err-ptb111}. It was also found that the difference
between the exact permittivity and the approximated one defined above, not
only is independent of $N$, but also of $\omega $. Thus a semi--infinite
periodic layered system can be simply modeled as a sequence of seven layers
(a central one plus three layers above and below this) in a constant
dispersionless permittivity background, the contribution of which to the
total permittivity is negligible. Because of the fast convergence in $\left|
p-q\right| ,$ calculations for periodic layered systems restricted to $N=3$
layers can already yield reasonably accurate optical constants.

\section*{Summary}

It was shown analytically and numerically that in polar geometry and normal
incidence the complex reflectivity coefficient of the right-- and
left--handed polarized light for periodic {\it (}layered{\it )} systems is
given by the Fresnel formula: only in the case of periodic (layered) systems
the $2\times 2$ matrix technique and the two--media approach become
equivalent, i.e., they lead to identical Kerr spectra. This in turn clearly
marks the limitations of the two--media approach, since applied to
inhomogeneous layered systems, the corresponding Kerr spectra are dominated
entirely by the optical activity of the surface layer.

Inter-- and intra--layer contributions to the optical conductivity for fcc Pt%
$_{N}$/Pt layered systems as calculated within the framework of the
spin--polarized relativistic screened Korringa--Kohn--Rostoker (SKKR) method
by means of a contour integration technique lead to optical constants, which
are in very good agreement with available experimental data for photon
energies $\omega >\delta $ ($\delta ${\it \ }being finite lifetime broadening%
{\it )}. The calculated optical constants for fcc Pt$_{N}$/Pt do depend on
the surface orientation; for $N\geq 3$ (number of Pt layers) they become
virtually independent of $N$.

The conclusion to be drawn for magnetic multilyer systems is rather simple:
only the $2\times 2$\ matrix technique that includes all multiple
reflections and optical interferences can produce reliable spectra for this
kind of systems. This in turn implies that a computational scheme such as
the SKKR method designed especially for systems with only two-dimensional
translational symmetry has to be applied exclusively in order to evaluate
the elements of the optical conductivity tensor.

\section*{Acknowledgments}

This work was supported by the Austrian Ministry of Science (Contract No.
45.451) and the Hungarian National Science Foundation (OTKA T030240,
T037856). The collaboration between L.S. and the Austrian partners 
(e.g., A.V.) was partially sponsored also by the Research and Technological
Cooperation Project between Austria and Hungary (Contract No. A-23/01).

%
%

%
\vfill
\newpage 
%
%
%

%
\vfill
\newpage 
%
%
%

%
\vfill
\newpage 
%
%

%
%

\end{document}